\documentclass[12pt]{article}
\usepackage{a4}
\usepackage{cite}
\usepackage{amsmath,amsfonts}
\usepackage{graphicx}

\newcommand{\msbar}{{\overline{MS}}}
\newcommand{\lmsbar}{\Lambda_{\overline{MS}}}
\newcommand{\beq}{\begin{eqnarray}}
\newcommand{\eeq}{\end{eqnarray}}
\newcommand{\op}{\operatorname}

\begin{document}
\title{The running of the coupling\\ in SU($N$) pure gauge theories}
\author{Biagio Lucini and Gregory Moraitis\\
\small{Physics Department, Swansea University, Singleton Park, Swansea SA2 8PP, UK}}
\date{}
\maketitle

\abstract{The running of the coupling is studied in SU(4) gauge theory using the Schr\"odinger functional technique. Up to energies of the order of the square root of the string tension $\sigma$, the running is found to agree with the two-loop perturbative formula. Relating the perturbative to the non-perturbative regime of the running and converting to the $\msbar$ scheme allows one to extract the ratio $\Lambda_{\msbar}/\sqrt{\sigma}$. The result is then used in combination with similar calculations present in the literature for SU(2) and SU(3) to extract $\Lambda_{\msbar}/\sqrt{\sigma}$ in the large $N$ limit. Our results for $N=3,4$ agree with a recent study of the same quantity performed using the Parisi mean field scheme as an intermediate scheme, while $\Lambda_{\msbar}/\sqrt{\sigma}$ in SU(2) turns out to differ by 2.5\%. Possible explanations of this discrepancy are discussed.}

\section{Introduction}
Asymptotic freedom~\cite{Gross:1973id,Politzer:1973fx} is one of the signatures of non-Abelian gauge theories. At short distances, the coupling is small and perturbation theory can be successfully used to compute observables. By dimensional transmutation, a scale is generated (of mass dimension 1) whose order of magnitude is dictated by the dynamics of the theory. In perturbation theory, it is convenient to associate with this scale the dimensionful multiplicative constant of the integrated perturbative beta function, which is called the $\Lambda$ parameter. The value of the $\Lambda$ parameter depends on the chosen perturbative scheme. Different regularisation schemes define different couplings. A good scheme is conventionally one for which low-order perturbative calculations work at energies close to the non-perturbative scale of the theory.

However good the scheme is, at long enough distances (low energies) the theory becomes inherently non-perturbative, and confinement sets in. In SU($N$) gauge theories (or more generally in theories with unbroken fundamental centre), this regime is characterised by the tension of the confining string, $\sigma$, which has dimensions of mass squared. Since SU($N$) gauge theories are theories with only one dynamically generated scale, it must be possible to relate $\Lambda$ and $\sigma$. For SU(2) and SU(3) gauge theory, this program has been successfully carried out by the Alpha collaboration~\cite{Luscher:1992zx,Luscher:1993gh,Capitani1998mq} using the Schr\"odinger functional (SF) scheme~\cite{Luscher:1991wu,Luscher:1992an} on a spacetime lattice. In the SF scheme, the theory is defined in a box of finite physical size, and the renormalised coupling is obtained through the effective action of the system with certain specified boundary conditions. This determines the coupling as a function of the extension of the box, $L$. On the lattice, simulations are carried out at different lattice spacings in order to extract continuum results. In doing so, the bare coupling must be carefully tuned for each lattice spacing to ensure that the physical size $L$ remains constant. An iterative procedure can then be set up to probe the theory over a large range of energies.

This technique has proved to be effective for exploring a range of couplings interpolating from the perturbative to the non-perturbative regime. On the perturbative end, the $\Lambda$ parameter in the SF scheme, $\Lambda_{\mathrm{SF}}$, can be determined using perturbation theory, in terms of the physical size of the box; on the non-perturbative end, the size of the box can be determined in terms of a non-perturbative quantity, e.g. the string tension $\sigma$. Since the relationship between the sizes of the box at the two ends is known, one can determine the ratio $\Lambda_{\mathrm{SF}}/\sqrt{\sigma}$. By calculating $\Lambda_{\mathrm{SF}}/\lmsbar$ using perturbation theory, the ratio can be re-expressed in the language of the more familiar $\msbar$ scheme as $\Lambda_{\mathrm{\msbar}}/\sqrt{\sigma}$.

In recent years, following renewed interest from string theory~\cite{Maldacena:1997re}, a program for non-perturbative studies of SU($N$) gauge theories in the large-$N$ limit~\cite{'tHooft:1973jz} has been developed (see e.g.~\cite{Lucini:2001ej,Lucini:2001nv,Lucini:2003zr,Lucini:2004yh,Lucini:2005vg,Bringoltz:2005rr,Bringoltz:2005xx,DelDebbio:2001sj,DelDebbio:2002xa,DelDebbio:2006df,DelDebbio:2007wk,Bali:2007kt}). A general conclusion is that observables in SU($N$) pure gauge theories have a smooth dependence on $N$ that, within a few percent, can be accounted for by a $1/N^2$ correction to the $N = \infty$ case for $N \ge 3$, and often also including the case $N = 2$. In this work, we investigate the dependence of $\lmsbar/\sqrt{\sigma}$ as a function of $N$ using the SF technique. To this end, first we formulate the problem of the running of the coupling for SU(N) gauge theory in the SF scheme~(Sect.~\ref{sect:2}) and numerically determine the value of $\lmsbar/\sqrt{\sigma}$ for SU(4) following the procedure described in~\cite{Luscher:1992zx,Luscher:1993gh} (Sect.~\ref{sect:3}). Then, using results available in the literature for SU(2) and SU(3) complemented with recent lattice determinations of the string tension, we discuss the behaviour of $\lmsbar/\sqrt{\sigma}$ as $N \to \infty$ (Sect.~\ref{sect:4}), comparing our results with those of~\cite{Allton:2008ty}, where the Parisi mean field improvement~\cite{Parisi:1980pe} is used as an intermediate scheme. A discussion of systematic errors follows (Sect.~\ref{sect:5}). Finally, we draw our conclusions in Sect.~\ref{sect:6}. A partial account of our calculation has already been published in~\cite{Lucini:2007sa}.
\section{The Schr\"odinger functional in SU($N$) gauge theory}
\label{sect:2}
%\subsection{Definitions}
%\label{sect:2.1}
Following Ref.~\cite{Luscher:1992an} (to which we refer for further details), we shall introduce the SF in SU($N$) lattice gauge theory. Consider two states ${\cal C}$ and ${\cal C}^{\prime}$ in the Schr\"odinger representation of a system described by the Hamiltonian $H$, whose associated action is given by $S$. The Schr\"odinger functional is the probability amplitude for ${\cal C}^{\prime}$ at time $t$ starting from ${\cal C}$ at time 0:
\beq
\label{eq:sfg}
Z[{\cal C},{\cal C}^{\prime}] = \langle {\cal C}^{\prime} | e^{- H t} | {\cal C} \rangle =  \int_{{\cal C},{\cal C}^{\prime}} D[\phi] e^{ - S[\phi]} \ ,
\eeq
where $\phi$ is the generic field configuration and $D[\phi]$ the measure of the path integral, which is taken at fixed boundary conditions. For a SU($N$) lattice gauge theory on a volume $V = L^4$ with $L = na$ ($n$ being an integer and $a$ the lattice spacing), described by the Wilson action, Eq.~(\ref{eq:sfg}) becomes
\beq
\mathcal{Z}[{\cal C},{\cal C}^{\prime}]=\int_{{\cal C},{\cal C}^{\prime}} D[U]e^{-S[U]} = e^{- \Gamma[{\cal C},{\cal C}^{\prime}]} \ ,
\eeq
where $\Gamma[{\cal C},{\cal C}^{\prime}]$ is the effective action of the system with the specified boundary conditions. The Wilson action appearing in the above definition is given by
\beq
\label{eq:wilsonact}
S[U]=\frac{1}{g_0^2}\sum_{p}\op{Tr}(1-U(p)) \ ,
\eeq
where $U(p)$ denotes the parallel transport of the link variables $U_{\mu}(x)$ ($x = (x^0,x^1,x^2,x^3)$ being a lattice point of integer coordinates and $\mu = 0,1,2,3$ the lattice directions) over the elementary square of the lattice ({\em plaquette}) $p$ and $g_0$ is the (bare) lattice coupling. The sum in Eq.~(\ref{eq:wilsonact}) must be taken over both orientations of the plaquettes. 

The boundary links $W$ are required to satisfy inhomogeneous Dirichlet boundary conditions for $k=1,2,3$,
\begin{equation}
\label{eq:links}
W_k(x)|_{x^0=0}=\exp\left(aC_k(x)\right),
\qquad W_k(x)|_{x^0=L}=\exp\left(aC'_k(x)\right),
\end{equation}
where $C_k$ and $C'_k$ are spatial boundary fields which need to be chosen.

Fixing the boundary field induces a background field in the bulk, and it is desirable to choose boundary fields which minimise the effect of the finite
lattice spacing. It was shown in~\cite{Luscher:1992an} that, for $N$ colours, the optimal choice are constant Abelian fields,
\begin{equation}
C_k=\frac{i}{L}
 \left( \begin{array}{cccc}
                       \phi_{k1}  & 0       & \cdots & 0 \\
                           0      & \phi_{k2}  & \cdots & 0  \\
                           \vdots & \vdots  & \ddots & \vdots \\
                           0      & 0       & \cdots & \phi_{kN}
                        \end{array} \right), \qquad
C'_k=\frac{i}{L}
 \left( \begin{array}{cccc}
                       \phi'_{k1}  & 0       & \cdots & 0 \\
                           0      & \phi'_{k2}  & \cdots & 0  \\
                           \vdots & \vdots  & \ddots & \vdots \\
                           0      & 0       & \cdots & \phi'_{kN}
                        \end{array} \right).
\end{equation}
Unitarity and stability considerations of the background field constrain the angles,
\begin{equation}
\label{eq:conditions}
\sum_{i=1}^{N}\phi_i=0,\qquad\phi_1<\phi_2<...<\phi_N,\qquad|\phi_i-\phi_j|<2\pi
\end{equation}
and similarly for $\phi'$ (from here on we drop the suffix $k$ on the angles and use the same choice for $k=1,2,3$). The ensemble of points satisfying the constraints~(\ref{eq:conditions}) is referred to as the fundamental domain.

The effective action can be written as an asymptotic series
\begin{equation}
\Gamma[B]=g_0^{-2}\Gamma_0[B]+\Gamma_1[B]+g_0^2\Gamma_2[B]+\dots \ .
\end{equation}
In this expansion, $\Gamma_0$ is the classical action, which can be evaluated analytically. The previous equation could be used directly to define a renormalised coupling via Monte Carlo simulations of the SF. However, the numerical determination of an effective action is a notoriously difficult problem. To get around this difficulty, one generally measures derivatives of the effective action. By introducing a dependence of the boundary links (and thus of the background field) on a real dimensionless parameter $\eta$, we can then define a renormalised coupling as
\begin{equation}
\label{eq:coupling}
\bar{g}^2=\frac{\Gamma'_0[B]}{\Gamma'[B]},\qquad\Gamma'[B]=\frac{\partial}{\partial\eta}\Gamma[B],
\end{equation}
for a particular choice of $\eta$.
%\subsection{Method}
%\label{sect:2.2}

The SF technique allows one to explore the running of the coupling for a wide range of energies, connecting the perturbative to the non-perturbative regime. On the lattice, for each energy scale $E_i$ probed, in order to extrapolate to the continuum limit, multiple simulations must be performed at different lattice spacings; hence, different bare couplings are needed. Those couplings (ordered from the one corresponding to the coarsest to the one corresponding to the finest simulated lattice spacing) are labeled sequentially by a second index $j$. Thus, we define $g_0^{i,j}$ as the $j$th bare coupling at the energy scale $E_i$, or, equivalently, the length scale $L_i = 1/E_i$. The running of the coupling is computed using a recursive procedure. We start by fixing the renormalised coupling for all $g_0^{0,j}$ in such a way that it is equal for the whole set, within errors. This common value of the renormalised coupling is called $\bar{g}^0$. Then, for each bare coupling $g_0^{0,j}$, the renormalised coupling is evaluated for boxes of size $2 L_0$ (by doubling the number of lattice sites in each direction) and the continuum limit is obtained by assuming a linear dependency in $a/L$ and extrapolating to $a \to 0$. We call the extrapolated value $\bar{g}^1$. Returning to a small lattice once again, a new set of couplings $g_0^{1,j}$ is then chosen in such a way that the value of the renormalised couplings for the size $L_1$ match $\bar{g}^1$. This ensures that the physical size of the box for the set $g_0^{1,j}$ is kept constant, and in particular equal to twice the size of the box corresponding to the set $g_0^{0,j}$. This procedure can then be iterated. If $L_0$ is the original size of the box, after $l$ iterations the size is $L_l=2^l L_0$. If the final set of couplings $g_0^{l,j}$ are in the asymptotic scaling regime of the theory, the product $L_l \sqrt{\sigma}$ can be obtained by determining $\sigma$ for each coupling in the set $g_0^{l,j}$. On the other end, if $\bar{g}^0$ is in the perturbative regime, $L_0$ can be obtained from the integrated two-loop beta function of the theory as
\beq
\label{eq:twoloop}
L_0=E_0^{-1}=\frac{1}{\Lambda_{SF}}\left(\frac{\beta_1}{\beta_0^2}+\frac{1}{\beta_0\bar g^2(E)}\right)^\frac{\beta_1}{2\beta_0^2}e^{-\frac{1}{2\beta_0\bar g^2(E)}} \ ,
\eeq
where 
\beq
\beta_0 = \frac{1}{(4\pi)^2}\frac{11}{3}N, \qquad \beta_1 = \frac{1}{(4\pi)^4}\frac{34}{3}N^2.
\eeq
Putting together the determination of $L_l \equiv L_{MAX}$ in terms of $\sqrt{\sigma}$ and $\Lambda_{SF}$, the value of $\Lambda_{SF}/\sqrt{\sigma}$ can be worked out.

In order to compare with other determinations, the SF scheme should be related to a more widely used scheme, e.g. the $\msbar$. Two regularisation schemes can be related by using first order perturbation theory. Consider two schemes, $A$ and $B$, in which the couplings $g_A$ and $g_B$ are defined. The corresponding $\Lambda$ parameters are found through the one-loop beta functions:
\beq
K = \frac{1}{\Lambda_A} e^{-\frac{1}{2 \beta_0 g_A^2}} \qquad \mbox{and}
\qquad K = \frac{1}{\Lambda_B} e^{-\frac{1}{2 \beta_0 g_B^2}} \ ,
\eeq
where $K$ is the length scale at which the coupling is measured. Putting together the two previous relationships yields
\beq
\frac{\Lambda_A}{\Lambda_B} = e^{- \frac{1}{2 \beta_0}\left(\frac{1}{g_A^2} - 
\frac{1}{g_B^2} \right)} \ .
\eeq
From a first order perturbative calculation of a physical quantity one gets
\beq
\frac{1}{g_A^2} - \frac{1}{g_B^2}  = k \ ,
\eeq
where $k$ is a constant depending on the details of the schemes. Once $k$ has been determined, we can rewrite the ratio of the $\Lambda$ parameters as
\beq
\frac{\Lambda_A}{\Lambda_B} = e^{- \frac{1}{2 \beta_0} k} \ .
\eeq

To relate the SF and $\overline{MS}$ scheme we can use the lattice scheme as an intermediate step, computing $\Lambda_{SF}/\Lambda_L$ and $\lmsbar/\Lambda_L$. The perturbative calculation relating the lattice and the SF schemes is described in~\cite{Luscher:1992an}, while the ratio
$\lmsbar/\Lambda_L$ for SU($N$) gauge theories (determined in Ref.~\cite{Dashen:1980vm}) is given by
\beq
\frac{\lmsbar}{\Lambda_L} = 38.85e^{- \frac{3 \pi^2}{11 N^2}} \ .
\eeq
Other perturbative calculations of $\lmsbar/\Lambda_L$~\cite{Hasenfratz:1980kn,Weisz:1980pu} give slightly discrepant results. However, the uncertainty coming from this perturbative calculation is some orders of magnitude smaller than the error on $\Lambda_{SF}/\Lambda_L$, and can safely be neglected.
\section{The running of the coupling in SU(4) gauge theory}
\label{sect:3}
\begin{figure}
\begin{center}
\includegraphics[scale=0.4]{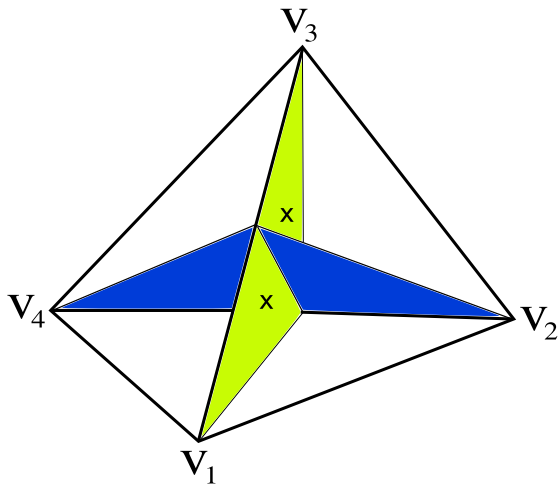}
\end{center}
\caption[The fundamental domain and its two planes of symmetry.]{The fundamental domain and its two planes of symmetry. The points marked $\mathbf x$ correspond to the angles~\eqref{eq:angles}.}
\label{fig:domain}
\end{figure}

%\subsection{Fundamental domain and simulation parameters}
%\label{sect:3.1}
We now specialise to SU(4). The fundamental domain can be described symmetrically by defining a one-to-one map between the set of angles $(\phi_1,\phi_2,\phi_3,\phi_4)$ and a point $\mathbf V$ in a certain bounded three-dimensional region,
\begin{equation}
\mathbf{V}=\tfrac{3}{4}(\phi_1\cdot\mathbf e_1+\phi_2\cdot\mathbf e_2+\phi_3
\cdot\mathbf e_3+\phi_4\cdot\mathbf e_4),\qquad\phi_i=\mathbf{V}\cdot
\mathbf e_i \ ,
\end{equation}
where $\mathbf e_i$ are the weights of the Lie algebra of SU(4) in the fundamental representation, normalised as $\mathbf e_i\cdot\mathbf e_j=\tfrac{1}{3}(4\delta_{ij}-1)$. With this normalisation, the vertices $\mathbf v_i$ of the fundamental domain are
\begin{equation}
\mathbf v_i=-\frac{3\pi}{2}\sum_{j=1}^{i}\mathbf e_j,\qquad i=1,2,3,4 \ ,
\end{equation}
describing a skewed tetrahedron (Figure \ref{fig:domain}).
One must then select, from inside the fundamental domain, the set of angles for each boundary, $C_k$ and $C'_k$, each set corresponding to a point inside the tetrahedron. These points will be members of a one-parameter family of angles parameterised by $\eta$, through which the renormalised coupling is defined by \eqref{eq:coupling}. In principle this choice is arbitrary and of no conceptual significance; however the signal-to-noise ratio of the Monte Carlo evaluation is highest when (i) the points are well away from the domain's edge, (ii) they are as far as possible from each other, and (iii) the two boundaries are on an equal footing. Geometrically, this corresponds to choosing two points related by a symmetry of the fundamental domain.

We choose the two points to be related by the symmetry reflecting about the plane through $\mathbf v_2$ and $\mathbf v_4$, and make the particular choice
\beq
\label{eq:angles}
\begin{array}{l}
\phi_1=-\tfrac{1}{2}\eta-\tfrac{1}{4}\pi\sqrt{2}\\     
\phi_2=-\tfrac{1}{2}\eta-\tfrac{1}{4}(2-\sqrt{2})\pi\\ 
\phi_3=\tfrac{1}{2}\eta+\tfrac{1}{4}(2-\sqrt{2})\pi\\
\phi_4=\tfrac{1}{2}\eta+\tfrac{1}{4}\pi\sqrt{2}
\end{array}
\qquad
\begin{array}{l}
\phi'_1=\tfrac{1}{2}\eta-\tfrac{1}{4}(2+\sqrt{2})\pi\\
\phi'_2=\tfrac{1}{2}\eta-\tfrac{1}{4}(4-\sqrt{2})\pi\\
\phi'_3=-\tfrac{1}{2}\eta+\tfrac{1}{4}(4-\sqrt{2})\pi\\
\phi'_4=-\tfrac{1}{2}\eta+\tfrac{1}{4}(2+\sqrt{2})\pi
\end{array}
\ .
\eeq
With this choice, we set $\eta=0$ and compute the renormalised coupling \eqref{eq:coupling} by calculating the expectation value of the observable
\begin{equation*}
\frac{\partial S}{\partial\eta}=-\frac{ia}{g_0^2L}\sum_{\mathbf x}\sum_{l=1}^3\left[(E_l(\mathbf x)+E'_l(\mathbf x))+(E_l(\mathbf x)+E'_l(\mathbf x))^\dagger\right],
\end{equation*}
\begin{equation}
\label{eq:observable}
E_l(\mathbf x)=\op{Tr}\left[cW_l(x)U_0(x+a\hat{l})U_l(x+a\hat0)^\dagger U_0(x)^\dagger\right]_{x^0=0},
\end{equation}
where $c=\op{diag}(-\tfrac{1}{2},-\tfrac{1}{2},\tfrac{1}{2},\tfrac{1}{2})$, and a similar expression holds for $E'(\mathbf x)$.
%\subsection{Algorithm details}
In our simulations, we have used the Cabibbo-Marinari algorithm, with one heat-bath update for every four over-relaxation steps. For the highest values of $\beta=8/g_0^2$ (i.e. smallest bare coupling), the number of configurations generated for each data point is of order $10^6$, and this number increases for decreasing $\beta$, up to $\sim10^7$ configurations for the lowest $\beta$.
%\subsection{Results}
\begin{table}[tbp]
\begin{center}
\begin{tabular}{|l|lll|l|}
 \hline
$\beta=8/g_0^2$  &  $L/a$ & $\bar{g}^2(L)$  & $\bar{g}^2(2L)$  & $\bar{g}^2(2L)_{a \to 0}$\\
 \hline
15.126	& 6	  & 1.0222(6)   & 1.2247(10) & \\
15.626	& 8	  & 1.0223(7)	& 1.2162(18) & 1.1892(11)\\
16.000  & 10	  & 1.0223(5)	& 1.2104(21) & \\
 \hline
14.137	& 6	  & 1.1893(3)   & 1.4833(13) & \\
14.632	& 8	  & 1.1892(4)   & 1.4705(19) & 1.4331(18)\\
15.007  & 10      & 1.1890(8)   & 1.4635(25) & \\
 \hline
13.142	& 6	  & 1.4329(3)   & 1.9043(14) & \\
13.631	& 8	  & 1.4332(4)	& 1.8798(24) & 1.8099(24)\\
14.000  & 10      & 1.4331(6)	& 1.8668(35) & \\
 \hline
12.190	& 6	  & 1.8098(7)   & 2.7198(38) & \\
12.668	& 8	  & 1.8102(6)	& 2.6548(49) & 2.4645(33)\\
13.030  & 10      & 1.8094(6)   & 2.6177(49) & \\
\hline
\end{tabular}
\end{center}
\caption{Pairs of renormalised couplings for fixed values of $\beta=8/g_0^2$.}
\label{tab:couplings}
\end{table}
\begin{figure}
\begin{center}
\includegraphics[scale=0.45]{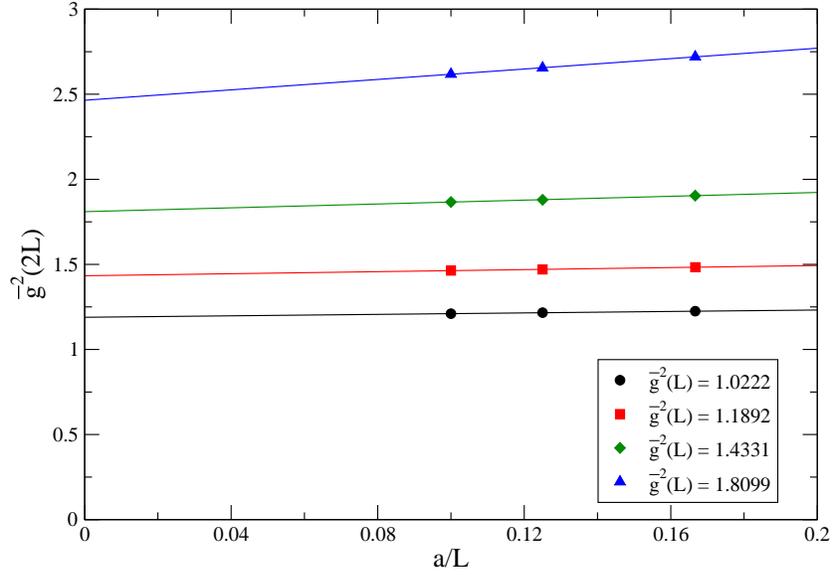}
\end{center}
\caption{Extrapolation to the continuum of the couplings in Tab.~\ref{tab:couplings}.}
\label{fig:g2cont}
\end{figure}

Following the outline of Sect.~\ref{sect:2}, we have performed four sets of simulations, each at three different lattice spacings. The results are shown in Tab.~\ref{tab:couplings}. Using the data in columns 2 and 4, we have extrapolated the renormalised coupling to the continuum using a linear fit in $a/L$, with a reduced $\chi^2$ of $\sim0.1$ for each fit. Although in principle higher order corrections in $a/L$ would need to be considered, the quality of the fit (showed by the low value of the $\chi^2$) suggests that the dominating contribution comes from the term proportional to $a/L$. The extrapolation is plotted in Fig.~\ref{fig:g2cont} and the result is shown in column 5 of Tab.~\ref{tab:couplings}. All the errors shown are statistical. Note, however, that there is an accumulation of errors due to the iterative nature of the procedure. This is because the statistical error in the extrapolated value of the renormalised coupling for some set $i$ (column 5) leads to an uncertainty on the value to which we tune the renormalised coupling in the next set $i+1$ (column 3). As in~\cite{Luscher:1992zx,Luscher:1993gh,Capitani1998mq}, it is convenient to interpret this mismatch as an error on the scale going from $L_i$ to $L_{i+1}$, and thus also from $L_0$ to $L_{MAX}$. The mismatch is small, so we can integrate the two-loop beta function to estimate the implied error on the scale at each stage, and then add the errors together in quadrature (since they come from independent sets of simulations). We find that $L_{MAX}=(2^4\pm0.04)L_0$.
 
\begin{table}[tbp]
\begin{center}
\begin{tabular}{|l|l|l|l|l|}
	\hline
$L/a$  & $\beta$ & $\bar{g}^2(L)$ & $a\sqrt\sigma$ & $L_{MAX}\sqrt\sigma$ \\ % & $\bar{g}2(2L)$  \\
	\hline
5 & 11.029  & 2.4644(4)  & 0.2093(10) & 1.046(5)\\
6 & 11.326	& 2.4646(7)  & 0.1611(10) & 0.966(6)\\
7 & 11.574	& 2.4645(18) & 0.1340(10) & 0.938(7)\\
8 & 11.782  & 2.4646(8)  & 0.1146(10) & 0.917(8)\\
 \hline
\end{tabular}
\end{center}
\caption{Data used for the extrapolation of $L_{MAX}\sqrt{\sigma}$ to zero lattice spacing for SU(4) gauge theory.}
\label{tab:bare}
\end{table}

\begin{figure}
\begin{center}
\includegraphics[scale=0.45]{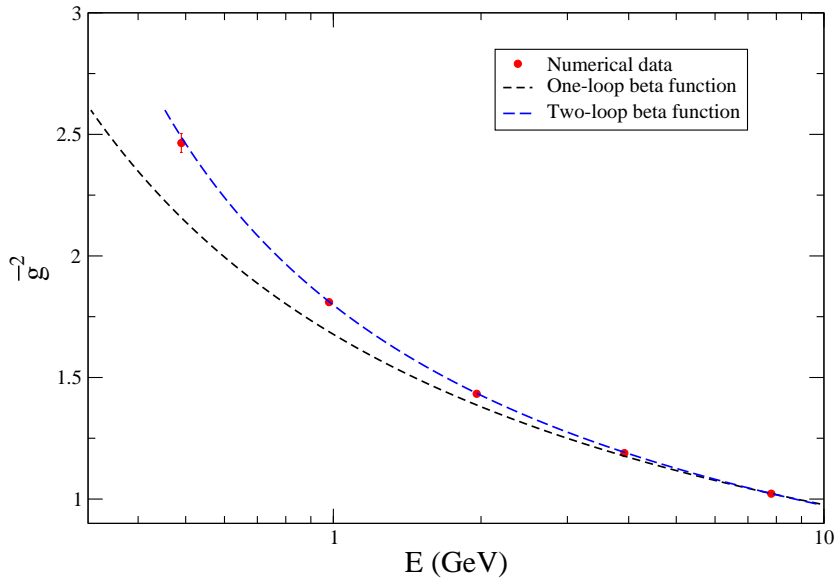}
\end{center}
\caption{Results of lattice simulations for the running coupling in the SU(4) theory, together with the one- and two-loop perturbative predictions. Energies in physical units are obtained by noting $\sqrt\sigma = 420$ MeV.}
\label{fig:coupling}
\end{figure}

At the most non-perturbative point (i.e. for $\bar{g}^2=2.4645$), we can express the scale in units of the string tension. In Tab.~\ref{tab:bare}, we quote the bare coupling at fixed renormalised coupling for different values of $L/a$, and the corresponding string tension $\sqrt\sigma$ in lattice units extracted using the interpolating formulae in~\cite{Lucini:2005vg} supplemented by newer data in~\cite{Allton:2008ty}\footnote{The results at $\beta = 11.574$ and $\beta = 11.782$ were obtained by extrapolating an unpublished string tension measurement at $\beta = 11.5$.}. The values of $L_{MAX}$ thus obtained then need to be extrapolated to the continuum. This requires a delicate fit, as it has been found (see e.g. the discussion in~\cite{Necco:2001xg}) that lattice corrections are large in both leading and next-to-leading order. We fit using
\begin{equation}
\label{eq:extrl0}
L_{MAX}\sqrt\sigma=L_{MAX}\sqrt\sigma|_{a=0}+c_1a\sqrt\sigma+c_2a^2\sigma \ ,
\end{equation}
and estimate the systematic error by varying the extremes of the fit and noting how much the extrapolated value shifts. This leads to a value of $L_{MAX}\sqrt\sigma=0.910(40)(100)$ where the first parentheses give the statistical error of the fit, and the second give our estimate of the systematic error. Using \eqref{eq:twoloop}, this gives
\begin{equation}
\Lambda_{SF}=0.253(10)(30)\sqrt\sigma=106.3\op{MeV}\pm4.2\op{MeV}\pm12.6\op{MeV} \ .
\end{equation}
where the value in MeV is obtained using $\sqrt\sigma = 420\op{MeV}$. Finally, we have calculated the ratio
\begin{equation}
\Lambda_{\msbar}/\Lambda_{SF}=2.08114(34)
\end{equation}
and, at the same time, calculated the improvement coefficient $c_t^{(1)}=-0.12005(15)$ (in the notation of Ref.~\cite{Luscher:1992an}). This immediately leads to
\begin{equation}
\Lambda_{\msbar}/\sqrt\sigma=0.527(21)(62) \ .
\end{equation}
We plot the renormalised coupling against the energy scale using the lattice data, together with the one- and two-loop perturbative predictions in Fig.~\ref{fig:coupling}. As was found in the previous studies of SU(2) and SU(3), two-loop perturbation theory gives excellent agreement with the data up to the scale of the string tension.

\begin{table}[tbp]
\begin{center}
\begin{tabular}{|c|c|c|c|}
  \hline
  $L/a$  & $\beta$ & $a\sqrt\sigma$ & $L_{MAX}\sqrt\sigma$ \\
  \hline
  5 & 2.5009 & 0.1835(10) & 0.918(5)\\
  6 & 2.5752 & 0.1438(10) & 0.863(6)\\
  7 & 2.6376 & 0.1179(10) & 0.825(7)\\
  8 & 2.6957 & 0.0984(10) & 0.787(8)\\
  9 & 2.7378 & 0.0859(10) & 0.773(9)\\
 10 & 2.7824 & 0.0751(10) & 0.751(10)\\
  \hline
\end{tabular}
\end{center}
\caption{Data used for the extrapolation of $L_{MAX}\sqrt{\sigma}$ to zero lattice spacing for SU(2) gauge theory. The $\beta$ values have been taken from~\cite{Luscher:1992zx} and the string tensions from~\cite{Lucini:2005vg,Allton:2008ty}.}
\label{Tab:l0su2}
\end{table}
\begin{table}[tbp]
\begin{center}
\begin{tabular}{|c|c|c|c|}
  \hline
  $L/a$  & $\beta$ & $a\sqrt\sigma$ & $L_{MAX}\sqrt\sigma$ \\
  \hline
  4 & 5.9044 & 0.2559(10) &	1.024(4) \\
  5 & 6.0829 & 0.1884(10) &	0.942(5) \\
  6 & 6.2204 & 0.1528(10) &	0.917(6) \\
  7 & 6.3443 & 0.1282(10) &	0.897(7) \\
  8 & 6.4527 & 0.1107(10) &	0.885(8) \\
  \hline
\end{tabular}
\end{center}
\caption{Data used for the extrapolation of $L_{MAX}\sqrt{\sigma}$ to zero lattice spacing for SU(3) gauge theory. The $\beta$ values have been taken from~\cite{Luscher:1993gh} and the string tensions from~\cite{Lucini:2005vg,Allton:2008ty}.}
\label{Tab:l0su3}
\end{table}

\section{Large-$N$ limit}
\label{sect:4}
Using the data for the SF from Refs.~\cite{Luscher:1992zx,Luscher:1993gh} and the data for the string tension from Refs.~\cite{Lucini:2005vg,Allton:2008ty}, we can compute the ratio $\Lambda_{\msbar}/\sqrt\sigma$ for SU(2) and SU(3) gauge theory in analogy with the SU(4) case discussed in the previous section. The only difference appears in the case of SU(3), where the simulations have been done with a Symanzik-improved action. Here we expect the linear coefficient in~\eqref{eq:extrl0} to be small (but not zero as the improvement is only calculated to 1-loop in perturbation theory), and indeed we observe this in our fit where we find $c_1=-0.20$ compared to $c_2=3.12$. We also fit with $c_1$ set to zero, and take the difference as an indication of the systematic error. For convenience, we report the data we have used for the extrapolations in Tabs.~\ref{Tab:l0su2}~and~\ref{Tab:l0su3}. In the continuum we find
\begin{equation}
L_{MAX}\sqrt\sigma= 0.603(17)(50)
\end{equation}
for SU(2) and 
\begin{equation}
L_{MAX}\sqrt\sigma= 0.854(3)(30)
\end{equation}
for SU(3). Using the conversion factors in Tab.~\ref{Tab:conversion}, those results imply
\begin{equation}
\Lambda_{\msbar}/\sqrt\sigma = 0.752(20)(60)
\end{equation}
and 
\begin{equation}
\Lambda_{\msbar}/\sqrt\sigma = 0.538(1)(20)
\end{equation}
for SU(2) and SU(3) respectively, where the first parenthesis is the statistical error and the second is an estimate of the systematic error coming from the uncertain form of the fit.
\begin{table}[tbp]
\begin{center}
\begin{tabular}{|c|c|}
  \hline
  $N$  & $\Lambda_{\msbar}/\Lambda_{SF}$\\
  \hline
  2 & 2.2446 \\
  3 & 2.0487 \\
  4 & 2.0811 \\
  \hline
\end{tabular}
\end{center}
\caption{Conversion coefficients from the $SF$ scheme to the $\msbar$ scheme.}
\label{Tab:conversion}
\end{table}

A reliable determination of the large-$N$ limit of $\Lambda_{\msbar}/\sqrt\sigma$ with the SF technique requires the determination of this quantity at larger values of $N$, which is outside the scope of this work. However, since a calculation in the large-$N$ limit of this quantity has been provided in~\cite{Allton:2008ty}, we can compare our data with the extrapolation reported there, to check the effects of possible systematic errors. We find that the results for $N=3,4$ are in good agreement, while for SU(2) the value of $\Lambda_{\msbar}/\sqrt\sigma$ obtained with the SF technique is higher (2.5\% when systematic errors of both calculations are included). However, there are indications that the SU(2) case is problematic in both calculations: with the SF technique, the extrapolation to the continuum limit proves to be less controlled than for SU(3) and SU(4) (an issue that can be resolved in principle by including larger volumes or using an improved action to determine the running of the coupling), while for the Parisi mean field method the plaquette in SU(2) might be affected by the end point of the bulk transition in the fundamental-adjoint plane (we refer to~\cite{Allton:2008ty} for further details). Although perfectly reasonable at this level, this discrepancy deserves further investigation. The agreement of the $N=3,4$ values computed with two independent methods shows that at large $N$ the systematic should be under control in both cases.
\section{Systematic errors}
\label{sect:5}
As we have seen in the previous section, albeit with larger errors, our results agree with those reported in~\cite{Allton:2008ty}. In order to better assess the scope of our findings, in this section we shall discuss in detail the systematic errors of our calculation.\\
\begin{figure}
\begin{center}
\includegraphics[scale=0.5]{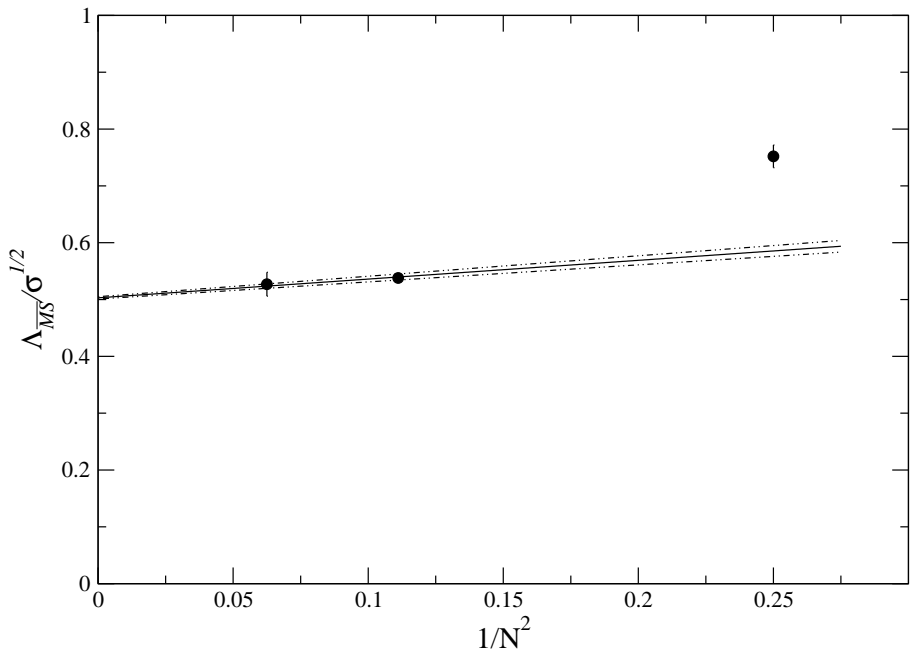}
\end{center}
\caption{The ratio $\Lambda_{\msbar}/\sqrt\sigma$ determined in this work as a function of $1/N^2$. The continuous line is the extrapolation to $N = \infty$ from Ref.~\cite{Allton:2008ty}, while the dot-dashed lines delimit the region at one sigma of confidence level (only the statistical errors are shown).}
\label{fig:largeNextr}
\end{figure}

The main sources of systematic errors are the following:
\begin{enumerate}
\item {\em Extrapolation of $L_{MAX} \sqrt\sigma$ to $a = 0$}. We have already mentioned that this is the biggest source of error in determining the $\Lambda$-parameter. In order to extrapolate $L_{MAX} \sqrt{\sigma}$ to $a = 0$, the string tension is measured for a number of lattice spacings and then the $a \to 0$ limit is taken according to Eq.~(\ref{eq:extrl0}). The difficulty is the largeness of both coefficients $c_1$ and $c_2$. We also have no information about the rest of the series. By changing the extremes of the fitting interval and comparing with fits including also cubic terms (where there are sufficient values of the string tension), we can measure the spread of those results to obtain a handle on the systematic error connected with the extrapolation. These rather large systematic errors could be improved if data for the string tension were available on finer lattices. We have systematics of 8\%, 4\%, 12\% for SU(2), SU(3) and SU(4) respectively.
\item{\em Interpolation of the string tension}. The error connected with the interpolation of the string tension can be easily evaluated by interpolating at values for which the string tension has been directly measured. Typically this error is also well below 1\%.
\item{\em Determination of the $\Lambda$ parameter using the two-loop beta function.} The $\Lambda$ parameter is determined by assuming the validity of the two-loop beta function at the most perturbative point. In principle we don't know at which energy scale the theory is well-described by two-loop perturbation theory. If the onset of the two-loop regime happens at lower energy scales, we could have used the second most perturbative point to determine $\Lambda_{SF}$. This procedure gives a systematic error of about 4\% for SU(3) and less than a percent for SU(4). On the other hand, it might be possible that two-loop perturbation theory fails even for our most perturbative point. To check that, we have used an approximation to the three-loop expression
\begin{eqnarray}
\label{eq:threeloop}
L_0=\frac{1}{\Lambda_{SF}}\left(\frac{\beta_1}{\beta_0^2}+\frac{1}{\beta_0\bar g^2(E)}\right)^\frac{\beta_1}{2\beta_0^2}e^{-\frac{1}{2\beta_0\bar g^2(E)}}
e^{-\frac{\beta_2^{SF}}{2 \beta_0^2}\bar g^2(E)}
 \ ,
\end{eqnarray}
where the scheme-dependent three-loop coefficient $\beta_2^{SF}$ has been determined in combination with $\Lambda_{SF}$ using the two most perturbative points. We found that the corresponding systematic error is of order 5\%. Starting at slightly lower coupling will reduce this error significantly.
\item {\em Scale uncertainty due to accumulation of iterative errors} This has been explained in Sect.~\ref{sect:3}, and contributes at under 1\%, so can safely be neglected.
\item {\em Large-$N$ extrapolation}. The large-$N$ extrapolation uses diagrammatic predictions truncated to leading correction in $1/N$. The truncation error and the onset of the large-$N$ regime can be determined respectively by adding higher order corrections and by excluding points at small $N$. Due to the fact that we have data only for three $N$ values, we do not have an estimate for these errors in our calculation.
\end{enumerate}
To sum up, an estimate of the total systematic error is 15\%, much greater than the statistical error, and almost all which comes from point 1 above.
\section{Conclusions}
\label{sect:6}
In this work, we have formulated the Schr\"odinger functional for SU(4) lattice gauge theory and studied it numerically. The resulting running of the coupling seems to be correctly described by two-loop perturbation theory down to energy scales of the order of $\sqrt\sigma$. This could be an indication of an underlying exact $\beta$ function~\cite{Ryttov:2007cx}, at least at large $N$~\cite{Armoni:2003fb}. Our results were also used to determine $\Lambda_{\overline{MS}}$. Using our calculation and the calculations for SU(2) and SU(3) given respectively in Refs.~\cite{Luscher:1992zx}~and~\cite{Luscher:1993gh}, we have performed a comparison with the extrapolation to $N \to \infty$ of~\cite{Allton:2008ty}, finding good agreement. Although the limited number of points in our study ($N$=2,3,4) does not allow us to perform a large-$N$ extrapolation, it is reassuring that the two studies give compatible results, as the dominant sources of systematic errors are different in the two cases. In order to perform a controlled extrapolation to $N=\infty$, we are currently extending the calculation to $N=6,8$. This will also require reducing the current systematic error on the value of $\Lambda_{\overline{MS}}$, which mostly comes from the extrapolation to the continuum limit of $L_{MAX}$. The smaller systematic error in SU(3) with respect to SU(2) and SU(4) suggests that this can be achieved by using an improved action that suppresses the linear term in $a \sqrt{\sigma}$. Finally, it would be interesting to compare the running of the coupling in the SF scheme to the running in the recently introduced interface tension scheme~\cite{deForcrand:2005pb,deForcrand:2005rg}.  
\section*{Acknowledgment}
We are indebted with M. L\"uscher and R. Sommer for sharing their insights on the SF with us. We also thank C. Allton, A. Armoni, P. Perez Rubio, F. Sannino and M. Teper for discussions on various aspects of this work. Numerical simulations have been performed on a Beowulf cluster partly funded by the Royal Society and STFC. B.L. is supported by a Royal Society University Research Fellowship. G.M. acknowledges financial support from a University of Wales Swansea Research Studentship. 
\bibliographystyle{myutcaps}
\bibliography{sf} %
\end{document}